\documentclass[conference]{IEEEtran}
\usepackage{amsmath,amssymb,amsfonts} 
\usepackage{amsthm}
\usepackage{cite}
\usepackage{graphicx}
\usepackage{textcomp}
\usepackage{booktabs}
\usepackage{tabularx}
\usepackage{makecell}
\usepackage{enumitem}
\usepackage{pifont}
\usepackage{xspace}
\usepackage{url}
\usepackage[table]{xcolor}
\usepackage[caption=false,font=footnotesize]{subfig}
\usepackage{algorithm}     
\usepackage[noend]{algpseudocode}
\usepackage{balance}
\usepackage{tikz}

\usepackage[export]{adjustbox}

\newcommand{\sysname}{\texttt{SwapLess}}

\newcounter{todonumber}
\stepcounter{todonumber}

\def\BibTeX{{\rm B\kern-.05em{\sc i\kern-.025em b}\kern-.08em
    T\kern-.1667em\lower.7ex\hbox{E}\kern-.125emX}}
\begin{document}

\author{
\IEEEauthorblockN{
Nathan Ng, 
Walid A. Hanafy, 
Prashanthi Kadambi\textsuperscript{\dag}, 
Balachandra Sunil,\\
Ayush Gupta, 
David Irwin, 
Yogesh Simmhan\textsuperscript{\dag}, 
Prashant Shenoy
}

\IEEEauthorblockA{
University of Massachusetts Amherst
}

\IEEEauthorblockA{
\textsuperscript{\dag}Indian Institute of Science
}
}

\title{
Collaborative Processing for Multi-Tenant Inference on Memory-Constrained Edge TPUs
}


\maketitle
\begin{tikzpicture}[remember picture,overlay]
  \node[anchor=north,inner sep=0pt,yshift=-0.35in]
    at (current page.north) {%
      \begin{minipage}{\textwidth}
        \small
        Published in the proceedings of IEEE DCOSS-IoT 2026\par
        \vspace{2pt}
        \hrule height 0.4pt
      \end{minipage}%
    };
\end{tikzpicture}

\begin{abstract}
IoT applications increasingly rely on on-device AI accelerators to ensure high performance, especially in low-connectivity and safety-critical scenarios. However, the limited on-chip memory of these accelerators forces inference runtimes to swap model segments between host and accelerator memory, incurring significant swapping overheads.
While collaborative processing by partitioning model execution across CPU and accelerator resources can reduce accelerator memory pressure and execution overhead, naive partitioning may worsen end-to-end latency by either shifting excessive computation to the CPU or failing to sufficiently reduce swapping, a problem that is further exacerbated in multi-tenant and dynamic environments.

To address these issues, we present \sysname, a system for adaptive, multi-tenant TPU–CPU collaborative inference on memory-constrained Edge TPUs. \sysname\ utilizes an analytic queueing model that captures partition-dependent CPU/TPU service times as well as inter- and intra-model swapping overheads across different workload mixes and request rates. Using this model, \sysname\ continuously adjusts both the partition point and CPU core allocation online to minimize end-to-end response time with low decision overhead. An implementation on Edge TPU-equipped platforms demonstrates that \sysname\ reduces mean latency by up to 63.8\% for single-tenant workloads and up to 77.4\% for multi-tenant workloads relative to the default Edge TPU compiler.

\end{abstract}

\section{Introduction}\label{sec:intro}
IoT applications increasingly rely on AI models to interpret and respond to their environments, such as smart cameras performing real-time activity recognition, wearables inferring health and mobility signals, and industrial sensors detecting anomalies for predictive maintenance~\cite{Wang2025:OnDeviceAISuvery}. While a common architectural approach is to offload computation from resource-constrained devices to nearby edge servers or the cloud, many IoT deployments cannot rely on offloading due to intermittent network connectivity or limited bandwidth~\cite{Liang2023, Ng2024}. Furthermore, in safety-critical and tactical scenarios such as disaster response and military operations, communication may be disrupted or restricted for security reasons~\cite{Abdelzaher2018:IoBT,Dong2025:Real-Offload}.
In such settings, IoT systems must execute inference locally, commonly referred to as on-device AI~\cite{Wang2025:OnDeviceAISuvery}.

To support the growing demands of AI models, IoT systems often rely on programmable AI accelerators, such as Google's Edge Tensor Processing Unit (Edge TPU) \cite{edgetpu2019} and Raspberry Pi AI HAT\cite{raspberrypi_ai_hat}, which enable low-end devices to perform tasks such as image classification and object detection with reasonable performance comparable to edge offloading.
Despite these advances, many accelerators have limited on-chip memory. 
For example, Edge TPUs provide only 8\,MB of SRAM for caching model parameters, which is insufficient for storing models such as InceptionV4 or ResNet50, resulting in high inference latency.

\begin{figure*}[t]
    \centering
    \begin{minipage}[b]{0.315\textwidth}
        \includegraphics[width=\textwidth]{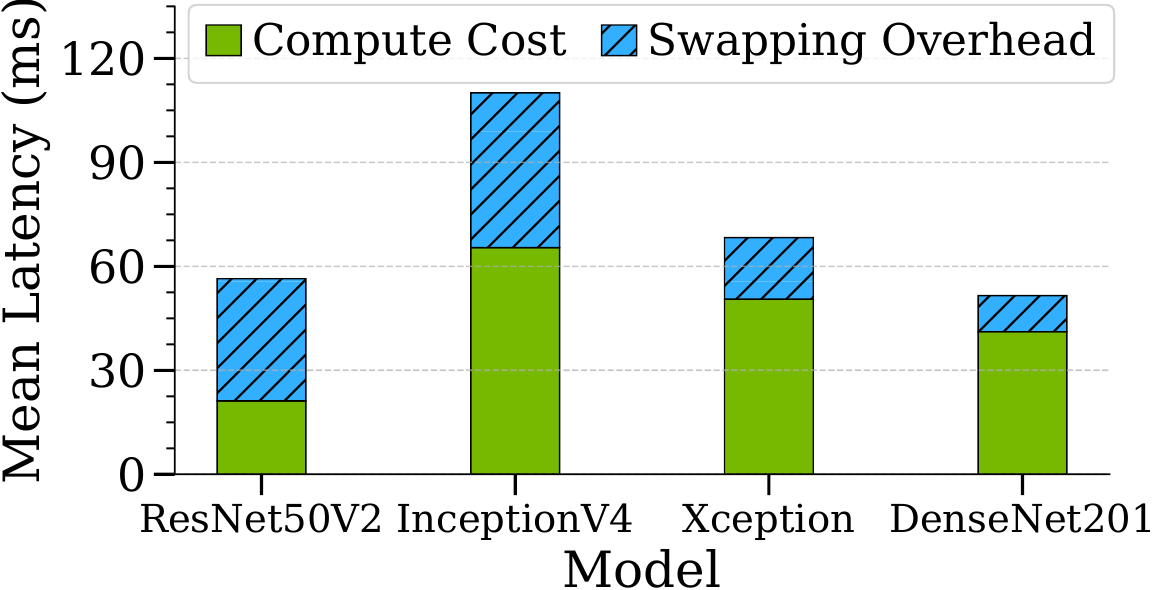}
        \caption{Intra-model swapping overhead can contribute up to 62.4\% of total TPU inference latency.}
        \label{fig:swap_overhead}
    \end{minipage}
    \hfill
    \begin{minipage}[b]{0.315\textwidth}
        \includegraphics[width=\textwidth]{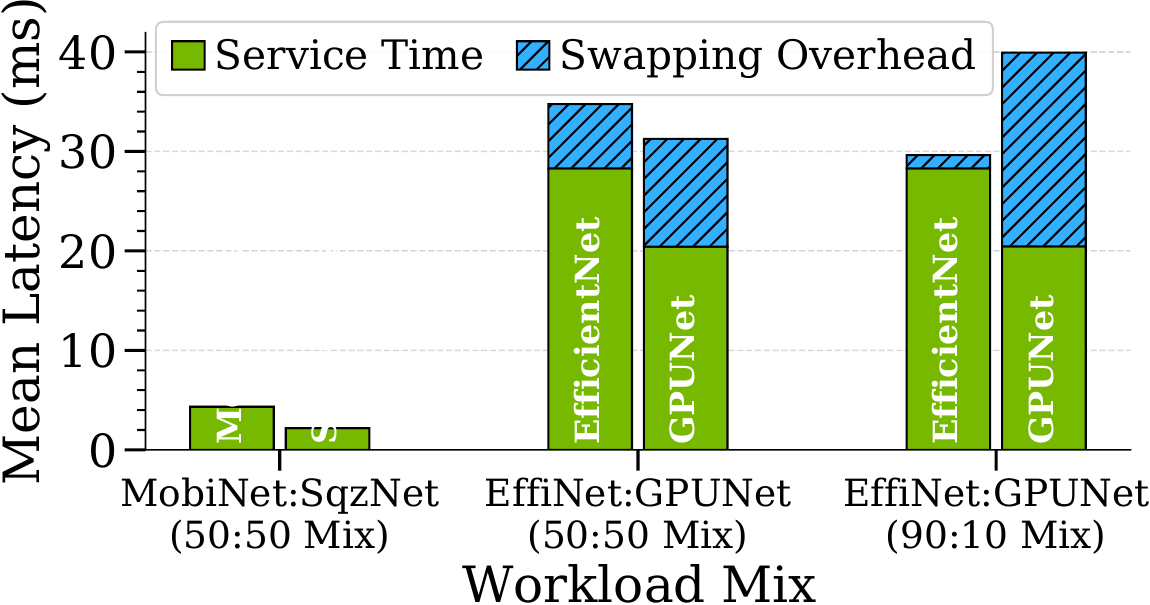}
        \caption{Inter-model swapping overhead can contribute up to 49\% of total latency.}
        \label{fig:multi_tenant_swap}
    \end{minipage}
    \hfill
    \begin{minipage}[b]{0.315\textwidth}
        \includegraphics[width=\textwidth]{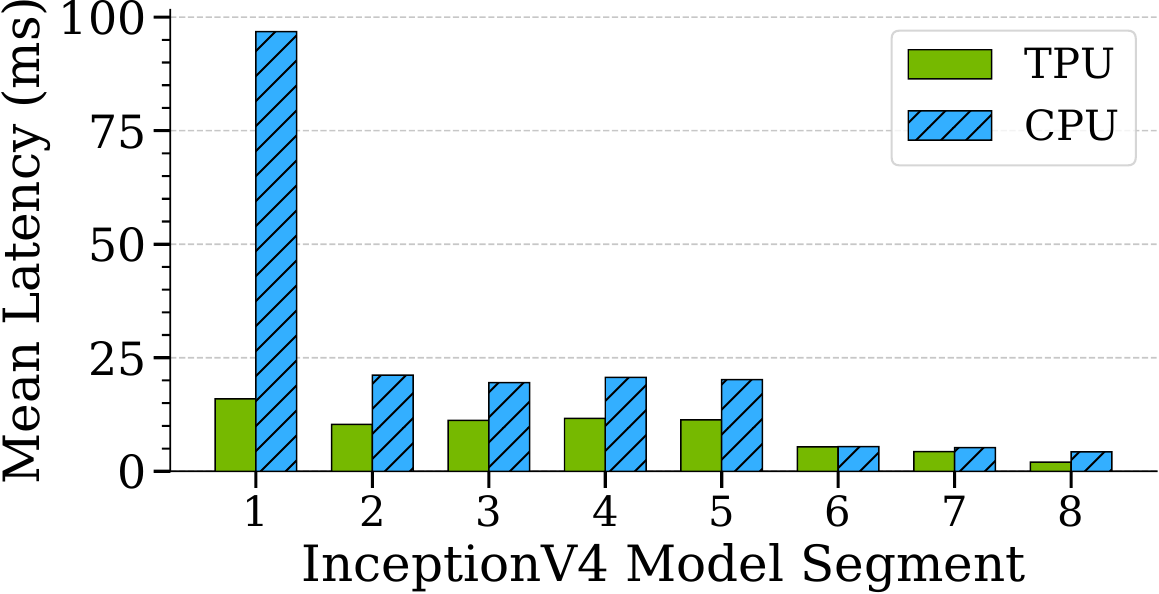}
        \caption{Comparable performance in later segments offers opportunities for collaborative inference.}
        \label{fig:inception_motivating}
    \end{minipage}
\end{figure*}

When a model exceeds TPU memory capacity, the inference runtime relies on memory-swapping approaches that sequentially swap model segments between host and accelerator memory during execution, introducing significant latency overhead. 
For instance, our analysis in Fig.~\ref{fig:swap_overhead} shows that swapping overhead can account for up to 62.4\% of the observed latency. 
This problem is further exacerbated in multi-tenant scenarios where multiple models run concurrently on the same device, increasing both the combined memory footprint and swap frequency.
To address this issue, prior research has focused on multi-TPU settings where models are split across TPU instances based on model size and request rates \cite{Zou2024:TPUpipeline, Yin2025:RESPECT, Villarrubia2023, Yin2022}.

Instead of scaling out to multiple accelerators, an alternative is to leverage the host CPU cores available on these edge platforms. Prior work has shown that, for the less-parallelizable layers of deep neural networks, general-purpose CPUs can deliver inference performance comparable to specialized accelerators, which often derive their advantage primarily from highly parallel layers~\cite{Li2022:GPU-CPU}. Building on this insight, we propose a collaborative processing approach that selectively offloads the trailing layers to the host CPU, providing a practical lever to reduce memory pressure and swapping overhead on the TPU.
However, as we show in Fig.~\ref{fig:compare_w_baseline}, naive partitioning may result in higher end-to-end latency than using the TPU alone. For example, choosing a partition that ensures the TPU-resident portion fits entirely in accelerator memory can substantially increase end-to-end latency by shifting too much computation to the CPU. Conversely, a conservative partition that offloads only the trailing layers may incur excessive swapping overhead.

To minimize end-to-end response time, a partitioning strategy must account for the relative processing speed of each segment on the CPU and TPU, the memory footprint of the TPU-resident segment and its associated swapping overhead, and the request arrival rate. 
These challenges are further exacerbated in multi-tenant, dynamic settings, where both the set of active models and their request rates can change over time. 
For example, the partitioning and request rate of one model directly affect the load on both the TPU and CPU, which in turn influence the queueing delays and swapping overheads experienced by other models.
This interdependence requires that offloading decisions and resource assignments be made holistically across all concurrently executing models, making model partitioning and CPU core allocation a complex, multi-dimensional resource allocation problem.

To address these challenges, we propose \sysname, a system for collaborative processing of multi-tenant inference on memory-constrained Edge TPU devices. 
\sysname\ continuously adjusts partitioning and CPU resource allocation online to minimize end-to-end response time with negligible decision overhead.
In designing, implementing, and evaluating \sysname, we make the following contributions: 
\begin{itemize}[leftmargin=*, nosep]
    \item At the core of \sysname\ decisions is an analytic queueing model that characterizes the performance of collaborative TPU-CPU inference in multi-tenant settings. The model jointly captures the effects of model partitioning, resource allocation, and inter- and intra-model swapping overheads across workload mixes and request rates. 
    \item We present the design of \sysname, a system that adaptively optimizes model partitioning and CPU core allocation on Edge TPU-equipped devices. \sysname\ uses a greedy hill-climbing resource allocation algorithm that leverages our analytic queueing model to minimize inference latency. 
    \item We implement and evaluate \sysname\ against three baselines across a range of workloads. Results show that \sysname\ reduces mean latency by up to 63.8\% for single-tenant workloads and up to 77.4\% for multi-tenant workloads compared to the default Edge TPU compiler.
\end{itemize}

\section{Background and Motivation}\label{sec:background}
This section provides background on Edge TPU inference, quantifies the memory swapping overhead in single- and multi-tenant scenarios, and discusses the opportunities and challenges of collaborative TPU-CPU inference.

\subsection{Edge TPU Hardware and Inference Execution}
Google’s Edge TPU is an application-specific integrated circuit (ASIC) with a systolic array architecture designed to accelerate AI inference on low-end devices. 
Each Edge TPU contains 8\,MB of on-chip SRAM for caching model parameters.  
An AI model is represented as a TensorFlow Lite computation graph and compiled with the Edge TPU compiler to generate optimized TPU-executable code. 
To run models larger than its on-chip memory, the TPU runtime first executes the model segment currently in memory, then sequentially swaps in the next segment from host memory and executes it, repeating until all segments are processed. 
In multi-tenant settings, the compiler prioritizes memory allocation based on the order in which models are specified in the compiler command, and some models may receive no initial memory allocation. 
For these models, execution requires first swapping their weights into memory, which incurs additional overhead compared to models whose weights are already resident on the TPU.
Since modern Transformer architectures involve dynamic attention, embedding layers, and non-linear operations that are not natively supported \cite{Reidy2023:TPUtransformer}, this work focuses on optimizing execution for standard convolutional models.

\subsection{Memory Swapping Overheads}
The Edge TPU’s limited memory poses a key performance bottleneck for many widely used model architectures, including InceptionV4, ResNet50V2, and DenseNet201, which exceed this limit. 
Fig.~\ref{fig:swap_overhead} illustrates this \emph{intra-model swapping} overhead using a micro-benchmark, where each model is partitioned into segments that fit in TPU memory. 
We profile the execution time of each segment and compare the total to the full model executed on the TPU, including swapping, to isolate the intra-model swapping overhead.
The results reveal that intra-model swapping overhead is a dominant factor in total inference latency, ranging from 20.2\% for DenseNet201 to 62.4\% for larger architectures such as InceptionV4.

This problem is further exacerbated in multi-tenant settings by \emph{inter-model swapping}, which occurs when the combined memory footprint of concurrent models exceeds the memory limit.
When requests for different models are executed in succession, parameters of the subsequent model that are not present in SRAM must be loaded from host memory, evicting existing weights and incurring swapping overhead before execution.
Fig.~\ref{fig:multi_tenant_swap} illustrates this overhead across different workload mixes.
In the workload mix with MobileNetV2 and SqueezeNet, no swapping occurs because their combined size fits within the memory.
However, models with larger combined footprints, such as EfficientNet and GPUNet, experience significant performance degradation. 
For a balanced 50:50 request mix, each request has roughly a 50\% chance of following a request for a different model, requiring an inter-model swap since its weights were evicted by the model used by the previous request. 
In this case, swapping overhead accounts for up to 35\% of total latency compared to standalone execution.
When the workload is skewed (e.g., a 90:10 mix), this penalty is even more pronounced for the less-frequent model, accounting for up to 49\% of its total execution latency. 

\subsection{Opportunities and Challenges of CPU Offloading}
Prior work~\cite{Li2022:GPU-CPU} has shown that GPUs and CPUs can achieve comparable performance in later, less parallelizable layers of AI models. 
Our experiments across a wide range of models confirm a similar opportunity on Edge TPUs. 
Fig.~\ref{fig:inception_motivating} compares the performance between TPU and CPU across segments of an InceptionV4 model. 
As shown, the first segment achieves a substantial TPU performance gain, which decreases in subsequent segments, with the last three segments showing comparable performance on TPU and CPU. 
We observe a similar trend across other models, and the results are omitted due to space constraints.  
This observation suggests an opportunity to minimize memory-swapping overhead and inference latency by reducing the TPU memory footprint through offloading model layers to the CPU.

Effectively capitalizing on this opportunity, however, requires careful resource allocation. 
First, the optimal partition point for each model must balance the relative processing speeds of the TPU and CPU for each segment, the memory footprint of the segments, and the interactions among co-located models, accounting for heterogeneous compute demands and varying memory-swapping overheads that can significantly impact overall performance. 
Second, in multi-tenant settings, the system must jointly consider model partitioning and CPU core allocation. 
The processing time of offloaded model segments on the CPU depends on both the number of cores allocated and the amount of computation offloaded, which in turn interacts with TPU memory allocation. 
Finally, runtime dynamics add further complexity, as request rates fluctuate and models may be added or removed over time. 
Consequently, resource allocation strategies must be adaptive and lightweight, responding to changing workloads without incurring excessive computational overhead.

\section{\sysname\ Design}\label{sec:design}
To address the above challenges, we propose \sysname, a system designed to optimize multi-tenant inference on Edge TPU-equipped devices. 
This section provides an overview of the \sysname\ design, introduces our proposed analytic queueing model that captures the effects of model partitioning, resource allocation, and memory swapping, and outlines its algorithm for joint model partitioning and core allocation.

\subsection{System Overview}
\begin{figure}[t]
    \centering
    \includegraphics[width=\linewidth]{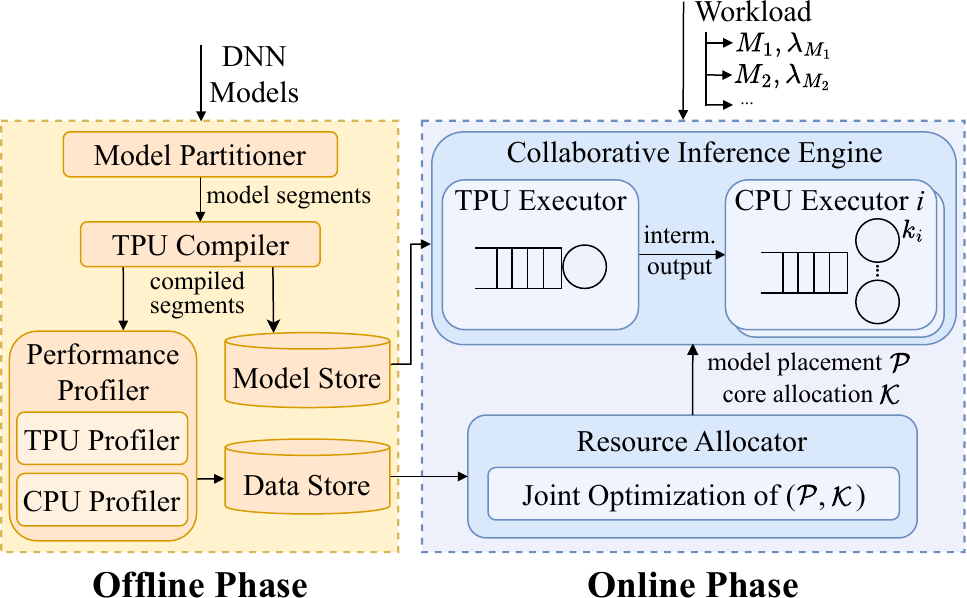}
    \caption{Overview of \sysname\ architecture. }
    \label{fig:system_overview}
\end{figure}

Fig.~\ref{fig:system_overview} presents the architecture of \sysname, which consists of an offline phase and an online phase.
In the offline phase, \sysname\ analyzes the AI models to identify viable partition points and partitions each model into a set of TPU prefixes and CPU suffixes.
It compiles the resulting segments for TPU execution and profiles them on both the TPU and CPU to characterize execution latency and resource usage.
The profile results and compiled binaries are stored for use at runtime.
During the online phase, the \sysname\ resource allocator uses this performance data and an analytic queueing model to jointly determine model partitioning and CPU core allocations, predicting system latency for candidate schemes to optimize end-to-end performance under memory and compute constraints. 
\sysname\ then executes requests according to the optimized allocation, assigning model prefixes to a global TPU worker and forwarding intermediate outputs to dedicated CPU executors, where suffixes are processed using the allocated CPU cores.

\subsection{Analytic Queueing Model}
To enable principled resource allocation, \sysname\ accounts for (i) queueing delays at the shared TPU under mixed workloads, (ii) CPU-side contention under per-model core allocations, and (iii) the latency impact of memory swapping.
\sysname\ uses an analytic model based on queueing theory, a well-established tool for modeling response times in computing systems, to estimate end-to-end latency under a given partitioning and resource allocation scheme.
These estimates are then used by \sysname's resource allocator to make adaptive resource allocation decisions.

\begin{table}[t]
\centering
\caption{Notations used in analytic models.}
\label{tab:notations}
\begin{tabularx}{\columnwidth}{@{} l X @{}}
\toprule
\textbf{Notation} & \textbf{Description} \\ \midrule
\rowcolor[gray]{.9} \multicolumn{2}{l}{\textit{Workload and Hardware Parameters}} \\
$\mathcal{M}$ & Set of $n$ AI models $\{M_1, \dots, M_n\}$. \\
$P_i$ & Total candidate partition points for $M_i$. \\
$M_{i, [a:b]}$ & Block of model $M_i$ from layer $a$ to $b$. \\
$d^{in}, d^{out}$ & Input and intermediate tensor size. \\
$B$ & Memory-to-TPU bandwidth. \\
$C$ & TPU SRAM capacity. \\
$K_{max}$ & Total available physical CPU cores. \\
\rowcolor[gray]{0.9} \multicolumn{2}{l}{\textit{Decision Variables}} \\
$p_i, k_i$ & Selected partition point and CPU cores allocated to $M_i$. \\
$\mathcal{P}, \mathcal{K}$ & Global model partitioning and core allocation vectors. \\
\rowcolor[gray]{0.9} \multicolumn{2}{l}{\textit{Performance Parameters}} \\
$\lambda_{M_i}$ & Request arrival rate for model $M_i$. \\
$\alpha_{M_i}(\mathcal{P})$ & TPU weight miss probability of $M_i$ under partitioning $\mathcal{P}$. \\
$T^{Load}$ & Latency to load weights into TPU memory. \\
$s^{TPU}, s^{CPU}$ & Service time for TPU prefix and CPU suffix. \\
$\mu$ & Service rate ($1/s$) \\
$\rho$ & Resource utilization ($\lambda/\mu$) \\
$E[W]$ & Expected queueing delay (TPU or CPU). \\
\bottomrule
\end{tabularx}
\end{table}
Consider a device that concurrently executes a set of AI models $\mathcal{M} = \{M_1, M_2, \dots, M_n\}$. 
Each model $M_i \in \mathcal{M}$ receives requests according to a Poisson process with arrival rate $\lambda_{M_i}$, where $\Lambda = \{\lambda_{M_1}, \dots, \lambda_{M_n}\}$ denotes the set of all arrival rates. 
Each model $M_i$ exposes $P_i$ candidate partition points that define potential boundaries between TPU and CPU execution. 
For a selected partition point $p_i \in \{0, \dots, P_i\}$, the prefix block $M_{i,[1:p_i]}$ executes on the Edge TPU, while the suffix block $M_{i,[p_i+1:P_i]}$ is offloaded to the host CPU. 
By convention, $p_i = 0$ corresponds to full CPU execution, and $p_i = P_i$ corresponds to full TPU execution. 

Let $s^{TPU}_{M_{i, [a:b]}}$ and $s^{CPU}_{M_{i, [a:b]}}$ denote the service time for a model block $M_{i, [a:b]}$ on the TPU and host CPU, respectively. 
For TPU execution, this service time includes the deterministic compute cost and the intra-model swapping overhead incurred when the prefix footprint exceeds TPU SRAM capacity $C$. 
To reduce the search space for real-time adaptation, we restrict each model to select a single partition point. 
We define $\mathcal{P} = \{p_1, \dots, p_n\}$ as the global model partitioning vector. 

To model inter-model swapping overheads, we introduce a weight miss probability $\alpha_{M_i}(\mathcal{P}) \in [0, 1]$, representing the likelihood that a request for $M_i$ requires a weight reload into TPU memory under global partitioning $\mathcal{P}$. 
As detailed in the following subsection, $\alpha_{M_i}$ is approximated by aggregate TPU memory footprint and relative arrival rates of concurrent models. 
We denote $T^{Load}_{M_{i, [1:p_i]}}$ as the inter-model swapping latency, which represents the time required to fetch the weights of the TPU prefix of model $M_i$ from host memory. 
This latency is a deterministic value derived from the size of block $M_{i, [1:p_i]}$ and the measured memory-to-TPU bandwidth $B$. 
Table~\ref{tab:notations} summarizes the notation used in our model. 

Given that the TPU processes multiple model segments in a first-come, first-served (FCFS) manner \cite{Liang2023}, we model the TPU as a single, unified $M/G/1/FCFS$ queueing system, where request arrivals follow a Poisson process and service times have a general (arbitrary) distribution. 
The expected wait time for such systems is well-known and is given by the Pollaczek-Khinchine formula \cite{Harchol2013}: 
\begin{equation} 
    E[W^{TPU}] = \frac{\lambda^{TPU} E[(s^{TPU})^2]}{2(1 - \rho^{TPU})} 
\end{equation}
where $\lambda^{TPU} = \sum_{i=1}^n \mathbb{I}(p_i > 0) \lambda_{M_i}$ is the aggregate arrival rate of model prefixes on the TPU, and $\rho^{TPU} = \lambda^{TPU} E[s^{TPU}]$ is the TPU utilization.
The effective service time $s^{TPU}(\mathcal{P})$ is a weighted average of the service times for all model prefixes under global partitioning $\mathcal{P}$, including the potential inter-model swapping latency:
\begin{equation}
    E[s^{TPU}(\mathcal{P})] = \sum_{M_i \in \mathcal{M}} \frac{\lambda_{M_i}}{\lambda^{TPU}} \left( \alpha_{M_i}(\mathcal{P}) T^{Load}_{M_{i, [1:p_i]}} + s^{TPU}_{M_{i, [1:p_i]}} \right)
\end{equation}

Following TPU processing, the suffix block $M_{i, [p_i+1:P_i]}$ is offloaded to the host CPU. 
To ensure performance isolation, \sysname\ partitions CPU resources by assigning each model $M_i$ a dedicated set of $k_i$ cores and an independent execution queue.  
Given that model service times are largely deterministic \cite{Liang2023}, we model the CPU suffix execution as an $M/D/k_i$ queueing system, where $D$ represents the deterministic service time distribution. 
The expected queueing delay $E[W^{CPU}_{M_i}]$ for model $M_i$ can be approximated as \cite{Shah:MDK}:
\begin{equation} 
    E[W^{CPU}_{M_i}] = \frac{1}{2} \left( \frac{1}{k_i\mu_{M_i}^{CPU} - \lambda_{M_i}} - \frac{1}{k_i\mu_{M_i}^{CPU}} \right) \label{eq:mdk_approx} 
\end{equation}
where the service rate $\mu_{M_i}^{CPU}$ is the inverse of the service time for the CPU suffix $s^{CPU}_{M_{i, [p_i+1:P_i]}}$. 

To model the data transfer costs between the TPU and host, let $d_{M_i}^{in}$ denote the input size of model $M_i$, $d_{p_i}^{out}$ denote the intermediate output size at partition point $p_i$. 
Further, let $\mathcal{K} = \{k_1, \dots, k_n\}$ denotes the core allocation for all models.
The expected end-to-end latency for model $M_i$ under a global configuration $(\mathcal{P}, \mathcal{K})$
\begin{equation}
    \begin{aligned}
        T_{M_i}^{e2e}(\mathcal{P}, \mathcal{K}) = & \; \mathbb{I}(p_i > 0) \left[ \frac{d_{M_i}^{in}}{B} + E[W^{TPU}] \right. \\
        & + \alpha_{M_i}(\mathcal{P}) T^{Load}_{M_{i, [1:p_i]}}  + s^{TPU}_{M_{i, [1:p_i]}} + \frac{d_{p_i}^{out}}{B} \Big] \\
        & + \mathbb{I}(p_i < P_i) \left[ E[W^{CPU}_{M_i}] + s^{CPU}_{M_{i, [p_i+1:P_i]}} \right] 
    \end{aligned}
\end{equation}

\noindent\textbf{Joint Partitioning and Core Allocation.}
Building on the latency model derived above, we formulate the optimization task as a Non-Linear Integer Program (NLIP). 
The objective is to minimize the aggregate weighted end-to-end latency across all models $\mathcal{M}$. 
Given a total physical capacity of $K_{\max}$ CPU cores, the optimization problem is defined as:
\begin{equation}
    \begin{aligned}
        \min_{\mathcal{P},\mathcal{K}} \quad
        & \sum_{M_i \in \mathcal{M}} \lambda_{M_i} \, T^{e2e}_{M_i}(\mathcal{P},\mathcal{K})
    \end{aligned}
\end{equation}
\text{s.t.} 
\begin{equation}
    p_i \in \{0, 1, \dots, P_i\}, \quad \forall M_i \in \mathcal{M}
    \label{c1}
\end{equation}
\begin{equation}  
    k_{i} \in \{0, 1, \dots, K_{\max}\}, \quad \forall M_i \in \mathcal{M}
    \label{c2}
\end{equation}
\begin{equation}    
    \mathbb{I}(p_i < P_i) \le k_i \le K_{\max} \cdot \mathbb{I}(p_i < P_i), \quad \forall M_i \in \mathcal{M}
\label{c3}
\end{equation}
\begin{equation}    
    \sum_{i=1}^{n} k_{i} \le K_{\max}
    \label{c4}
\end{equation}
where (\ref{c1}) and (\ref{c2}) set the discrete boundaries for partition points and core allocations. 
(\ref{c3}) ensures that at least one core is allocated to any model with a CPU suffix and prevents CPU allocation for full-TPU execution. 
(\ref{c4}) enforces the global physical core capacity $K_{\max}$.

\noindent\textbf{Modeling the Weight Miss Probability. }
To model the weight miss probability $\alpha_{M_i}$ of model $M_i$, let $W(\mathcal{P}) = \sum_{M_i \in \mathcal{M}} \text{size}(M_{i, [1:p_i]})$ be the aggregate TPU memory footprint under partitioning $\mathcal{P}$. 
Since memory eviction policies are typically proprietary and undocumented, we adopt a conservative approximation of $\alpha_{M_i}$ under two operational regimes:
\begin{equation} 
    \alpha_{M_i} = \begin{cases} 0 &  W(\mathcal{P}) \leq C \text{ or }  |\mathcal{P}| = 1
    \\ 1 - \frac{\lambda_{M_i}}{\lambda^{TPU}} & \text{otherwise} \end{cases} 
    \label{eq:alpha}
\end{equation}
In the first regime, we set $\alpha_{M_i}$ to zero because the cumulative memory requirement $W(\mathcal{P})$ fits within TPU capacity $C$.
Following a cold start, the system reaches a steady state where weights persist on-chip.
Furthermore, for single-tenant execution ($|\mathcal{P}| = 1$), we empirically observe that the reload probability remains zero regardless of model size, as the TPU runtime maintains weight persistence across inferences by fetching only required tiles rather than performing a full context reload.
The second regime applies when the aggregate footprint $W(\mathcal{P})$ exceeds $C$. Here, the TPU behaves as a shared-occupancy cache where $M_i$ is subject to eviction by competing requests of other models. 
We approximate the residency of $M_i$ using the stationary probability that at least one intervening request for a different model has displaced its weights since its last execution. 
The term $1 - \frac{\lambda_{M_i}}{\lambda^{TPU}}$ serves as a conservative upper bound, assuming that any intervening request for a different model results in an eviction of $M_i$.

\subsection{Greedy Hill-Climbing Resource Allocation}
Building on the response-time estimates from the analytic model, the optimization problem is inherently nonlinear due to the nonconvexity of the queueing models.
While such problems can be addressed using standard non-linear solvers, the computational overhead is prohibitive for the resource-constrained devices targeted in this work. 
Therefore, we propose a lightweight heuristic that jointly optimizes model partitioning and core allocation, detailed in Algorithm~\ref{alg:greedy_search}.
\begin{algorithm}
\caption{Greedy Hill-Climbing Resource Allocation}
\label{alg:greedy_search}
\begin{algorithmic}[1]
\small
\Require Models $\mathcal{M}$, Rates $\Lambda$, Max Cores $K_{\max}$, TPU Capacity $C$
\Ensure Optimized Partitioning $\mathcal{P}$, Core Allocation $\mathcal{K}$
\State $\mathcal{P} \gets [p_1, \dots, p_n] = [0, \dots, 0]$ 
\State $\mathcal{K} \gets \textsc{PropAlloc}(\mathcal{M}, \Lambda, \mathcal{P}, K_{\max})$
\State $L_{curr} \gets \sum_{M_i \in \mathcal{M}} \lambda_{M_i} \, T_{M_i}^{e2e}(\mathcal{P}, \mathcal{K})$
\While{True}
    \State $L'_{m,h} \gets \infty, \mathcal{K}'_{m,h} \gets \text{null}, \forall M_m \in \mathcal{M}, \forall h \in \{1, 2\}$
    \For{each $M_m \in \mathcal{M}$}
        \For{$h \gets 1$ to $2$}
            \If{$p_m + h \leq P_m$}
                \State $\mathcal{P}' \gets [p_1, \dots, p_m+h, \dots, p_n]$ 
                \State $\mathcal{K}'_{m,h} \gets \textsc{PropAlloc}(\mathcal{M}, \Lambda, \mathcal{P}', K_{\max})$
                \State $L'_{m,h} \gets \sum_{M_i \in \mathcal{M}} \lambda_{M_i} \cdot T_{M_i}^{e2e}(\mathcal{P}', \mathcal{K}'_{m,h})$ 
            \EndIf
        \EndFor
    \EndFor
    \State $L_{min} \gets \min_{m,h}(L'_{m,h})$
    \If{$L_{min} < L_{curr}$}
        \State $(m^*, h^*) \gets \arg\min_{m,h} L'_{m,h}$
        \State $p_{m^*} \gets p_{m^*} + h^*$; \quad $L_{curr} \gets L_{min}$; \quad $\mathcal{K} \gets \mathcal{K}'_{m^*, h^*}$ 
    \Else
        \State \textbf{break}
    \EndIf
\EndWhile
\State \Return $\mathcal{P}, \mathcal{K}$
\end{algorithmic}
\end{algorithm}
The algorithm first assigns all model to the CPU, with each model's core allocation proportional to its CPU workload, and computes the expected system latency using the analytic model (Lines 1–3).
In each iteration, it considers moving up to two layers to the TPU for each model $M_m$ (Lines 6–9), performs a proportional fair-share allocation that assigns a larger integer share of the $K_{\max}$ cores to models with higher CPU workloads (Line 10), and estimates the resulting system latency $L'_{m,h}$ (Line 11).
The search follows a greedy policy, committing the model and step $(m^*, h^*)$ that most reduces system latency (Lines 12–16).
This process repeats until no further moves reduce the expected system latency. 
The runtime overhead is minimal ($<$ 3\,ms on the tested platform) since the algorithm relies solely on closed-form computations in the analytic models.

\section{Implementation}\label{sec:imp}
We implement \sysname\ using TensorFlow Lite, a widely deployed runtime for edge AI applications.
\sysname\ follows a decoupled design with an offline profiling phase and an online orchestration engine.
In the offline phase, \sysname\ takes pre-trained TensorFlow frozen graphs as input and performs a topological traversal to identify candidate partition points that separate the graph along a single edge.
\sysname\ then partitions the graphs at each candidate point into a TPU prefix and a CPU suffix using the GraphSurgeon library \cite{onnx_graphsurgeon}. 
Segments are then converted to 8-bit quantized format via the TensorFlow Lite Converter~\cite{tflite_converter}, and TPU prefixes are compiled with the TPU Compiler~\cite{coral_edgetpu_compiler} to generate optimized binaries.
Finally, \sysname\ performs a one-time profiling of all segments to obtain service times for its analytic model.
Note that other service-time prediction methods such as \cite{Yao2018} can be seamlessly integrated into \sysname.

During the online phase, \sysname\ continuously monitors request rates using a sliding window and periodically runs its resource allocation algorithm to update partition points and CPU core assignments.
Requests are executed via a global TPU worker that maintains an FCFS queue for prefix execution.
Intermediate outputs are forwarded to model-specific CPU threadpools for suffix execution, with pool sizes determined by the allocation scheme.

\section{Experimental Evaluation}\label{sec:eval}
In this section, we present our experimental setup, validate our analytic queueing model, and compare \sysname\ against three baselines across workloads and utilization levels.

\subsection{Experimental Setup}\label{sec:eval_setup}
\subsubsection{Hardware}
Our evaluation is conducted on a Google Coral USB Accelerator featuring an Edge TPU coprocessor capable of performing 4 trillion operations per second (TOPS). 
The Edge TPU connects to a Raspberry Pi\,5 via USB 3.0, which features a quad-core ARM Cortex-A76 CPU at 2.4\,GHz and 8\,GB of LPDDR4X SDRAM.

\subsubsection{Models}
We evaluate \sysname\ using nine representative AI models spanning a diverse range of computational requirements and memory footprints. 
Their key characteristics are summarized in Table \ref{tab:model_characteristics}.  
\begin{table}[t]
\caption{Characteristics of evaluated AI models.}
\label{tab:model_characteristics}
\centering
\resizebox{.86\linewidth}{!}{%
\begin{tabular}{lccc}
\toprule
\textbf{Model} & \textbf{Size (MB)} & \textbf{FLOPs (G)} & \textbf{\# Partition Points} \\ 
\midrule
SqueezeNet          & 1.4                & 0.81               & 2                    \\
MobileNetV2         & 4.1               & 0.30               & 5                    \\
EfficientNet        & 6.7               & 0.39               & 6                    \\
MnasNet             & 7.1               & 0.31               & 7                    \\
GPUNet              & 12.2               & 0.62               & 5                    \\
DenseNet201         & 19.7               & 4.32               & 7                   \\
ResNet50V2          & 25.3               & 4.49               & 8                    \\
Xception            & 26.1              & 8.38               & 11                    \\
InceptionV4         & 43.2              & 12.27              & 11                    \\ \hline
\end{tabular}
}
\end{table}



\subsubsection{Baselines}
We use three representative baselines.
\begin{itemize}
    \item \textit{Edge TPU Compiler} \cite{coral_edgetpu_compiler}: This baseline represents the current industry deployment standard. It employs a static co-compilation strategy where multiple models are compiled together to share the TPU memory. Models may incur inter-model swapping overhead during inference.
    \item \textit{Threshold-based Partitioning}: This heuristic evaluates layers starting from the last one and offloads a layer to the CPU if its CPU execution time is within 10\% of its TPU execution time.
    It focuses on per-model hardware efficiency and does not account for queueing effects or multi-tenant resource contention.
    \item \textit{\sysname\ ($\alpha=0$)}: To isolate the impact of inter-model swapping, this variant of \sysname\ excludes the weight miss probability $\alpha_{M_i}(\mathcal{P})$. It relies on queueing models but assumes no swapping ($\alpha = 0$), thereby ignoring memory-swapping latency.
\end{itemize}

\noindent Note that \textit{Threshold-based Partitioning} and \textit{\sysname\ ($\alpha=0$)} are baselines introduced in this work, as prior work does not provide directly comparable baselines for our single-Edge-TPU device setting.

\subsection{Model Validation}
First, we validate the proposed analytic queueing model across single- and multi-tenant deployment scenarios.

\subsubsection{Single-tenant deployment validation}
\begin{figure}[t]
    \centering
    \subfloat[]{
        \includegraphics[width=0.472\columnwidth]{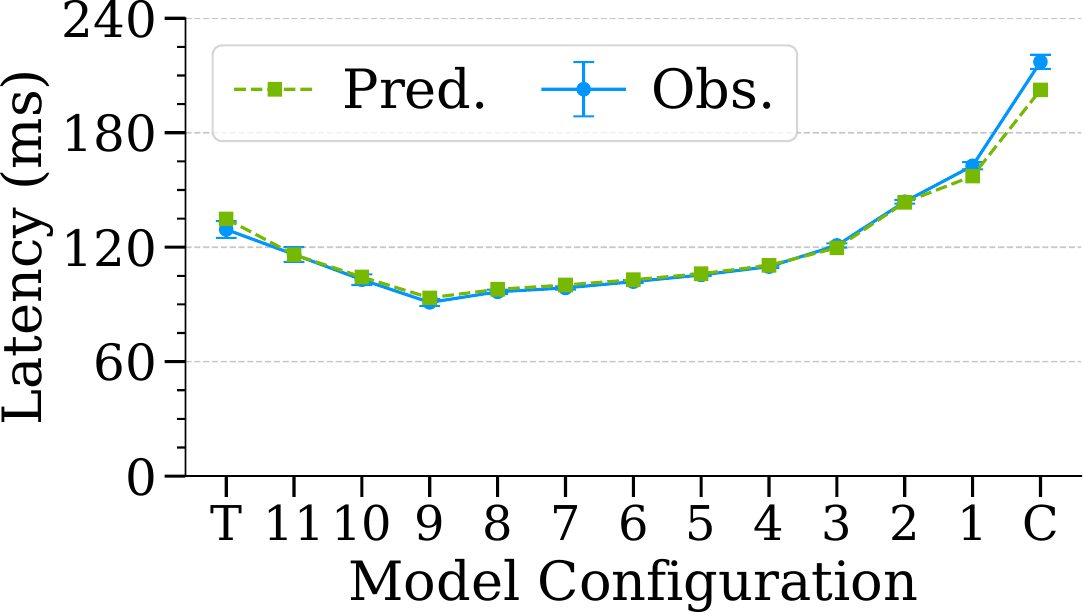}
        \label{fig:single_exec_strategy}
    }
    \subfloat[]{
        \includegraphics[width=0.472\columnwidth]{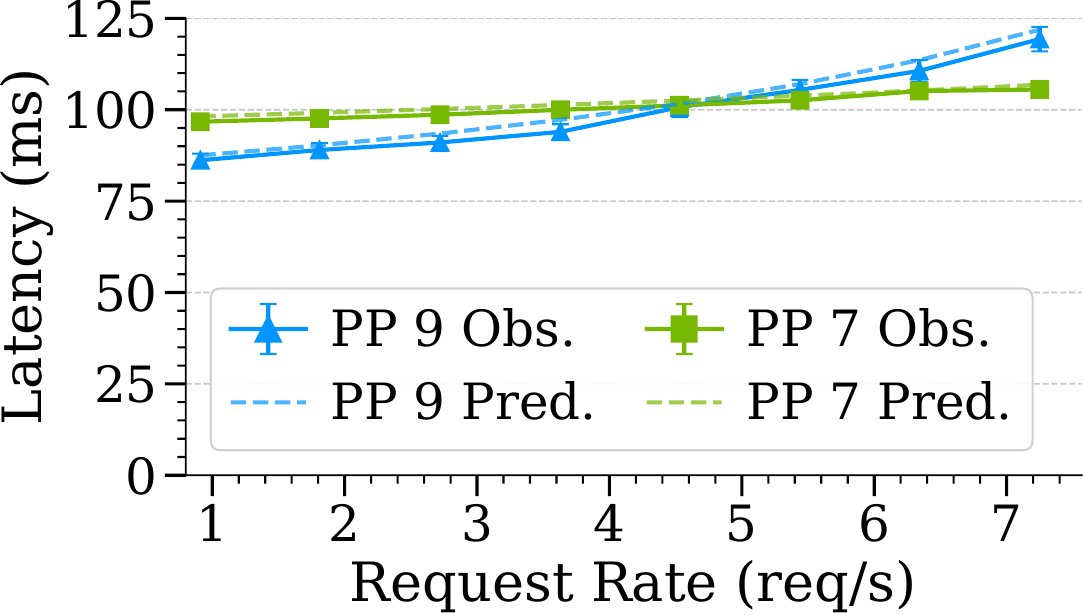}
        \label{fig:single_rps_scaling}
    }
    \caption{Model validation for single AI model deployments: 
    (a) InceptionV4 performance under different partition points (PPs). 
    (b) InceptionV4 performance across request rates.}
    \label{fig:single_model_validation}
\end{figure}

Fig.~\ref{fig:single_model_validation} shows the accuracy of \sysname’s analytic model using InceptionV4 under low load ($\rho=0.2$) across partition configurations. 
As shown, \sysname\ closely matches observed execution times across configurations, achieving a mean absolute percentage error (MAPE) of 1.9\%, with 92.3\% of predictions within $\pm5\%$ and all within $\pm10\%$ of the observed mean latency.
Fig.~\ref{fig:single_rps_scaling} validates the prediction accuracy across different request rates.
Importantly, the optimal partition configuration depends on the request rate: PP~9 yields lower average latency below 4.5~RPS, while PP~7 becomes optimal above 4.5~RPS. 
This highlights the inefficiency of static configurations and motivates the need for our analytic model. 
Similar accuracy is observed for other models and results are omitted due to space constraints.

\begin{figure*}[t]
    \centering

    \subfloat[]{
        \includegraphics[width=0.29\textwidth]{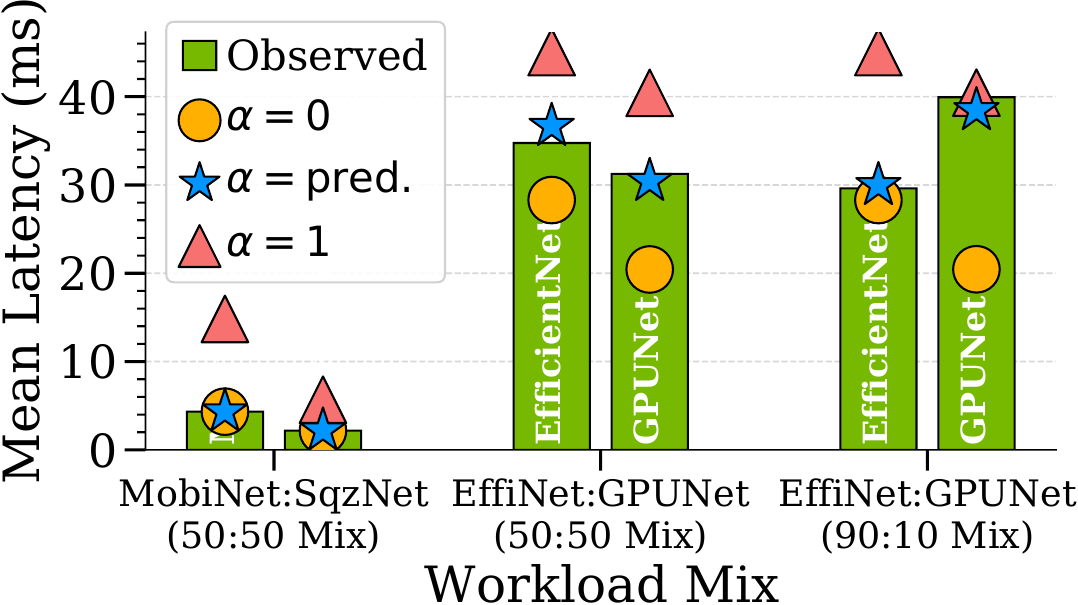}
        \label{fig:model_loading_overhead}
    }
    \hfill
    \subfloat[]{
        \includegraphics[width=0.29\textwidth]{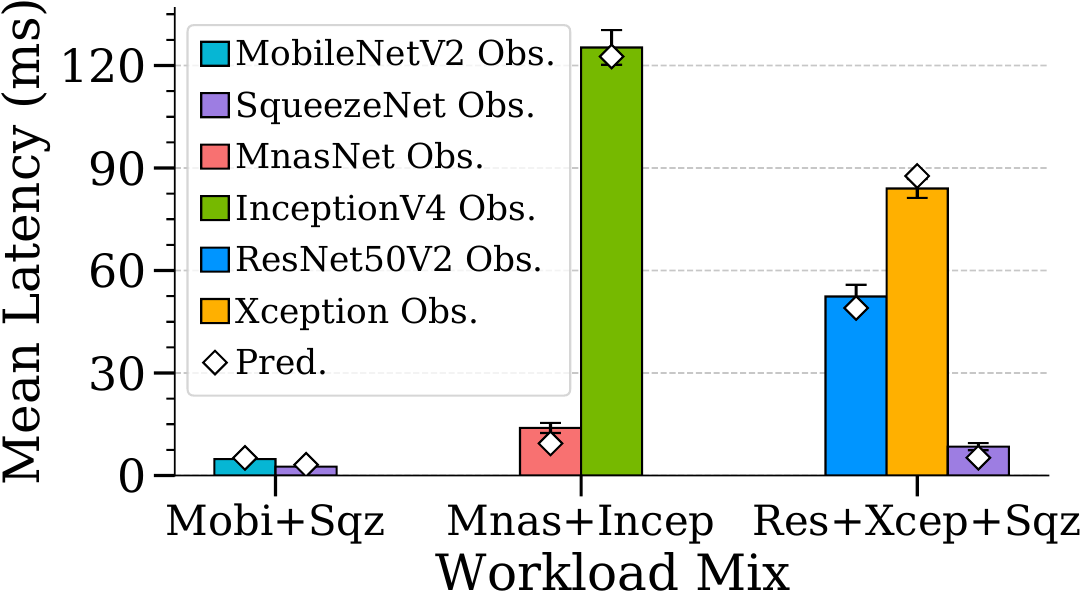}
        \label{fig:multi_model_combinations}
    }
    \hfill
    \subfloat[]{
        \includegraphics[width=0.29\textwidth]{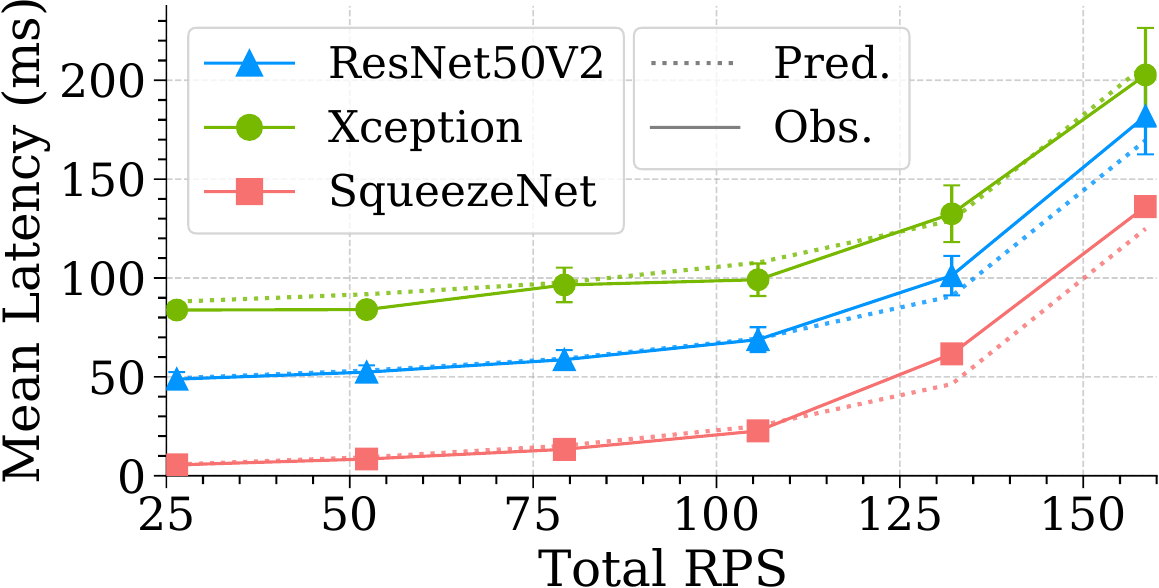}
        \label{fig:multi_model_rps}
    }
    \vspace{-.2em}
    \caption{Model validation for multi-tenant deployments: 
    (a) Validation of the $\alpha$ parameter across multiple workload mixes.
    (b) Accuracy across different concurrently running model combinations. 
    (c) Accuracy across request rates for a model combination.}
    \label{fig:multi_model_validation}
\end{figure*}
\subsubsection{Multi-tenant deployment validation} 
Fig.~\ref{fig:multi_model_validation} shows the accuracy of \sysname's analytic model across diverse model mixes. Fig.~\ref{fig:model_loading_overhead} demonstrates that the weight miss probability $\alpha$ effectively characterizes model-loading overhead.
In the first scenario, MobileNetV2 and SqueezeNet together fit within the Edge TPU memory and do not incur swapping overhead, which is correctly captured by the first regime in (\ref{eq:alpha}) with $\alpha = 0$. 
In the second scenario, EfficientNet and GPUNet exceed the TPU memory capacity. 
Here, \sysname\ sets $\alpha = 0.5$ for each model based on a 50:50 request mix, accurately modeling inference performance. 
In the third scenario, with a 90:10 request skew, \sysname\ adjusts $\alpha$ for each model based on the request distribution, providing precise latency predictions with a MAPE of 2.2\%. 
Fig.~\ref{fig:multi_model_combinations} compares observed and predicted latencies across different model mixes, while Fig.~\ref{fig:multi_model_rps} validates \sysname's analytic model under different request rates. 
For each workload mix, each model’s request rate is configured to generate an equal TPU load. 
The results show that \sysname's predicted latencies closely match the observed mean latencies for all workload mixes with a MAPE of 6.8\%, confirming the accuracy of the analytic model.

\noindent\textbf{Key takeaway.} \textit{\sysname\ accurately predicts end-to-end latency across partition configurations, request rates, and workload mixes, achieving a MAPE of 1.9\% for single-tenant and 6.8\% for multi-tenant workloads.}

\subsection{Comparison with Baselines}
\begin{figure*}[t]
    \centering
    \subfloat[$\rho = 0.2$]{
        \includegraphics[width=.985\textwidth]{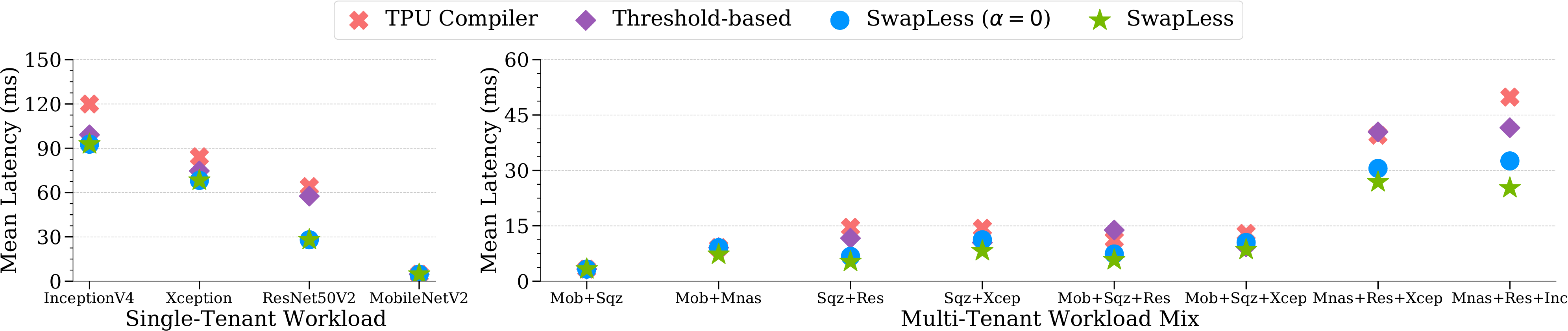}
        \label{fig:baseline_02}
    } \vspace{.5em}
    \subfloat[$\rho = 0.5$]{
        \includegraphics[width=.985\textwidth]{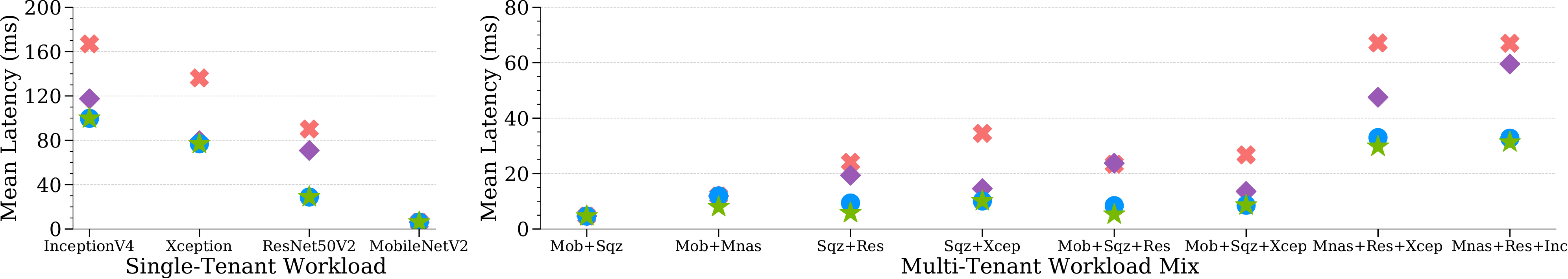}
        \label{fig:baseline_05}
    }
    \vspace{-.2em}
    \caption{Latency comparison between \sysname\ and baselines under different TPU utilization $\rho$.}
    \label{fig:compare_w_baseline}
\end{figure*}
Fig.~\ref{fig:compare_w_baseline} compares the performance of \sysname\ with representative baselines across diverse model mixes, varying TPU utilization levels ($\rho$), and both single-tenant (left) and multi-tenant (right) workloads.
For multi-tenant workloads, each model contributes equally to the total request load, and the y-axis reports the mean latency across all models.
As seen, when all models fit within the TPU memory (e.g., MobileNetV2 and MobileNetV2+SqueezeNet), all approaches achieve similar performance as executing the workload on TPU does not incur swapping overhead. 
Conversely, for workloads whose memory footprint exceeds TPU capacity, \sysname\ demonstrates significant advantages. 
Under low utilization ($\rho = 0.2$), \sysname\ reduces mean latency by up to 56.2\% in single-tenant settings and 68.0\% in multi-tenant scenarios relative to the TPU Compiler. 
These substantial gains are primarily attributed to \sysname’s explicit modeling of memory swapping, which enables more effective resource allocation under memory pressure. 
It is worth noting that in single-tenant environments, setting $\alpha=0$ yields performance identical to that of the full \sysname\ since no inter-model swapping occurs.
As load increases to a moderate level ($\rho = 0.5$), the benefits of \sysname\ become even more pronounced as it accounts for both queueing delays and memory swapping overheads, achieving latency reductions of up to 63.8\% and 77.4\% in single- and multi-tenant settings, respectively. 
In contrast, threshold-based partitioning performs worse than the TPU compiler for some workloads (e.g., MobileNetV2+SqueezeNet+ResNet), as it does not account for queueing delays and swapping overheads.
Overall, \sysname\  consistently matches or outperforms all baselines, achieving the lowest overall latency.

\noindent\textbf{Key takeaway.} \textit{\sysname\ reduces end-to-end latency by up to 63.8\% and 77.4\% compared to the baselines across both single- and multi-tenant deployments.}

\subsection{Impact of Dynamic Workloads}
\label{sec:eval_dynamic}
\begin{figure}
    \centering
    \includegraphics[width=.913\linewidth]{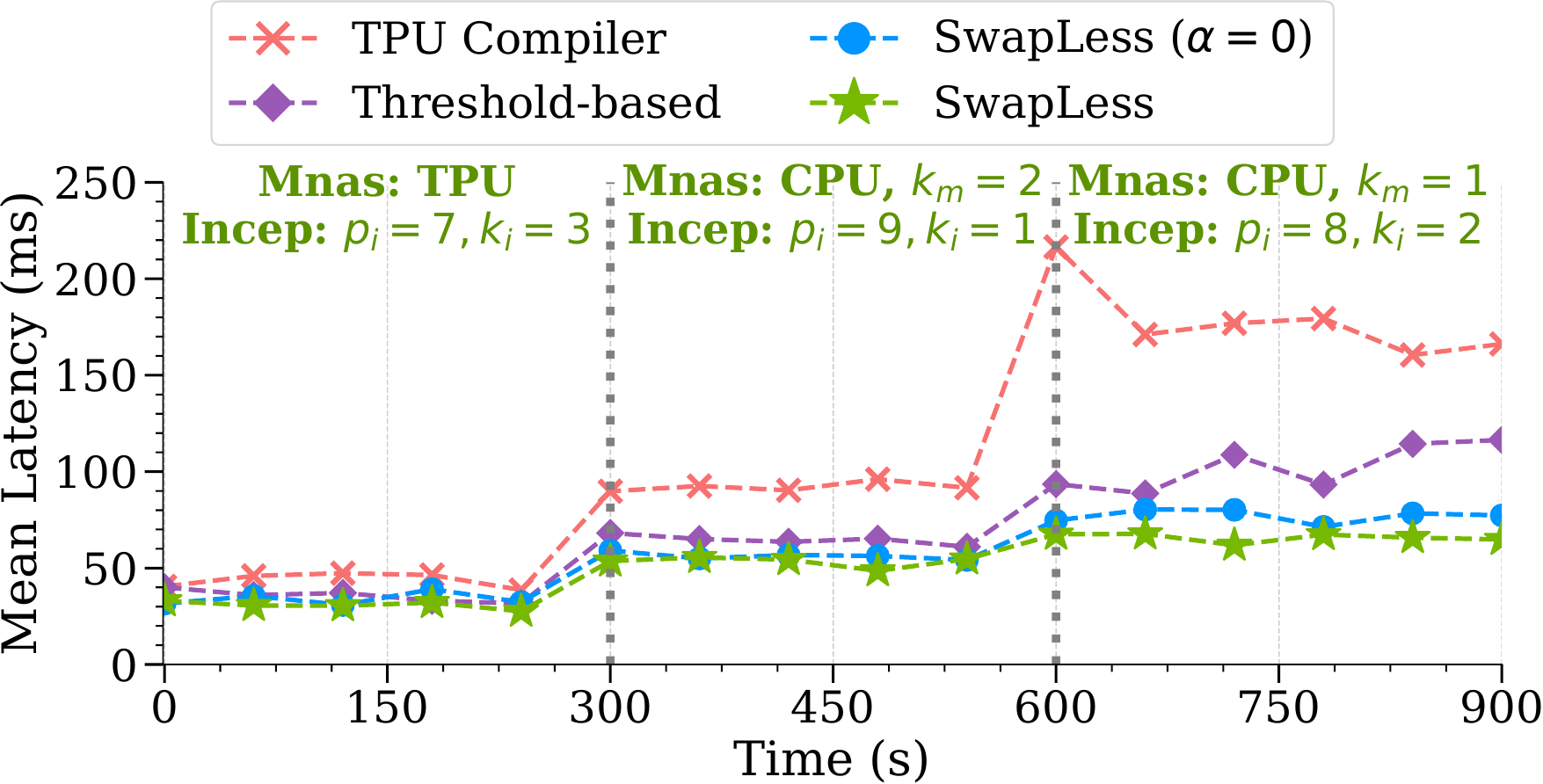}
    \vspace{-.75em}
    \caption{Performance under dynamic request rates for MnasNet and InceptionV4. Request rates are initially (5,1) RPS, then (5,3) from 300–600\,s, and (5,5) from 600–900\,s. 
    }
    \label{fig:dynamic_exp}
\end{figure}
We now evaluate \sysname’s ability to sustain low latency under time-varying load.
Fig.~\ref{fig:dynamic_exp} shows the performance of \sysname\ and representative baselines as workload request rates vary over time.
To enable fast reconfiguration, \sysname\ preloads a small set of representative model partitions for low-overhead configuration switching, with further optimization of the switching mechanism left for future work.
Initially, \sysname\ partitions TPU memory to accommodate both models, with MnasNet fully running on the TPU and InceptionV4 partitioned at $p_i = 7$.
As InceptionV4’s request rate increases to 3~RPS, \sysname\ dynamically offloads the MnasNet model to the CPU, reallocating the freed TPU resources to InceptionV4 to handle the increased load.
When the load further increases to 5~RPS, \sysname\ offloads one additional InceptionV4 layer to CPU to mitigate TPU queueing delays and reallocates an additional core to handle its increased CPU load.
The overhead of the adaptive algorithm is negligible and only incurs less than 3\,ms per invocation.
Through adaptive resource allocation, \sysname\ achieves up to 75.1\% latency reduction compared to baselines. 

\noindent\textbf{Key takeaway.} \textit{\sysname\ adapts to workload changes with minimal overhead, reducing mean latency by up to 75.1\%.}

\section{Related Work}
\noindent\textbf{Edge TPU Inference Optimization. }
Several work has attempted to address the memory constraints on Edge TPUs \cite{coral_pipeline, Yin2022, Villarrubia2023, Zou2024:TPUpipeline, Yin2025:RESPECT, Sun2025:SAPar}.  
For example, RESPECT \cite{Yin2025:RESPECT} proposes a reinforcement learning–based DNN scheduler for pipelined Edge TPUs. 
SAPar \cite{Sun2025:SAPar} introduces a surrogate-assisted partitioning tool to distribute DNN segments across multiple Edge TPUs to reduce latency.
However, these approaches rely on multiple TPU devices and do not address the challenges of single-device execution, which is the focus of our work.

\noindent\textbf{Cross‑processor Inference.} Prior work has explored partitioning DNNs across processors such as CPUs and GPUs to improve inference performance\cite{Kim2019,Jia2022,Ling2023,Jeong2022,Sen2025}. 
For example, Band \cite{Jeong2022} partitions and schedules model subgraphs across mobile processors, while BlastNet~\cite{Ling2023} partitions DNNs into paired CPU-GPU blocks across processors.
However, they assume a unified memory architecture in which accelerators can directly access host memory without incurring swapping overhead, and thus do not address the challenges introduced by memory‑constrained accelerators.

\noindent\textbf{Multi-tenant Inference.}
Another line of work focuses on coordinating and optimizing the execution of multiple concurrent DNNs \cite{Jiang2018,Fang2018,Han2021,Zhang2023}.
For instance, LegoDNN \cite{Han2021} uses block-grained scaling to dynamically combine reusable blocks at runtime, optimizing accuracy and latency.  
POS \cite{Zhang2023} employs a reinforcement learning–based operator scheduler to coordinate concurrent DNN execution to reduce latency. 
However, these approaches generally assume sufficient memory and computing resources, and do not handle the challenges introduced by memory-constrained accelerators.
\section{Conclusion}
This paper presented \sysname, a system for multi-tenant TPU–CPU collaborative inference on memory-constrained Edge TPU devices. 
\sysname\ uses an analytic queueing model that captures collaborative inference performance characteristics for adaptive resource allocation. 
Evaluation results show that \sysname\ reduces mean inference latency by up to 77.4\% compared to the default Edge TPU compiler.

\section*{Acknowledgment}
This research is supported by NSF grants 2211302, 2211888, 2213636, 2105494, 23091241, 2325956, and the Army Research Laboratory under Cooperative Agreement W911NF-17-2-0196 (IoBT CRA). 

\bibliography{main, DNNbibs}

@misc{edgetpu2019,
  author       = {Google},
  title        = {{Edge TPU: Run Inference at the Edge}},
  year         = {2019},
  howpublished          = {\url{https://www.coral.ai/docs/edgetpu/inference/}},
  note         = {Accessed: 2026-01-23}
}

@book{Harchol2013,
  author = {Harchol-Balter, M.},
  title = {Performance Modeling and Design of Computer Systems: Queueing Theory in Action},
  publisher = {Cambridge University Press},
  year = {2013},
  chapter = {14},
}

@inproceedings{Yao2018,
author = {Yao, Shuochao and others},
title = {{FastDeepIoT: Towards Understanding and Optimizing Neural Network Execution Time on Mobile and Embedded Devices}},
year = {2018},
isbn = {9781450359528},
url = {https://doi.org/10.1145/3274783.3274840},
doi = {10.1145/3274783.3274840},
numpages = {14},
booktitle = {SenSys'18}
}

@article{Liang2023,
author = {Liang, Qianlin and others},
title = {{Model-driven Cluster Resource Management for AI Workloads in Edge Clouds}},
year = {2023},
issue_date = {March 2023},
publisher = {Association for Computing Machinery},
address = {New York, NY, USA},
volume = {18},
number = {1},
issn = {1556-4665},
url = {https://doi.org/10.1145/3582080},
doi = {10.1145/3582080},
journal = {ACM Trans. Auton. Adapt. Syst.},
month = mar,
articleno = {2},
numpages = {26}
}

@INPROCEEDINGS{Ng2024,
  author={Ng, Nathan and others},
  booktitle={MILCOM'24}, 
  title={{Collaborative Inference in Resource-Constrained Edge Networks: Challenges and Opportunities}}, 
  year={2024},
  volume={},
  number={},
  doi={10.1109/MILCOM61039.2024.10773876}}

@inproceedings {Jiang2018,
author = {Angela H. Jiang and others},
title = {Mainstream: Dynamic {Stem-Sharing} for {Multi-Tenant} Video Processing},
booktitle = {USENIX ATC'18},
year = {2018},
isbn = {978-1-931971-44-7},
pages = {29--42},
url = {https://www.usenix.org/conference/atc18/presentation/jiang},

month = jul
}

@inproceedings{Fang2018,
author = {Fang, Biyi and others},
title = {{NestDNN: Resource-Aware Multi-Tenant On-Device Deep Learning for Continuous Mobile Vision}},
year = {2018},
isbn = {9781450359030},
url = {https://doi.org/10.1145/3241539.3241559},
doi = {10.1145/3241539.3241559},
numpages = {13},
location = {New Delhi, India},
booktitle = {MobiCom'18}
}

@inproceedings{Han2021,
author = {Han, Rui and others},
title = {{LegoDNN: block-grained scaling of deep neural networks for mobile vision}},
year = {2021},
isbn = {9781450383424},
url = {https://doi.org/10.1145/3447993.3483249},
doi = {10.1145/3447993.3483249},
pages = {406–419},
numpages = {14},
location = {New Orleans, Louisiana},
booktitle = {MobiCom'21}
}

@inproceedings{Zhang2023,
author = {Zhang, Ziyang and others},
title = {{POS: An Operator Scheduling Framework for Multi-model Inference on Edge Intelligent Computing}},
year = {2023},
isbn = {9798400701184},
url = {https://doi.org/10.1145/3583120.3586953},
doi = {10.1145/3583120.3586953},
location = {San Antonio, TX, USA},
booktitle = {IPSN'23}
}

@inproceedings{Kim2019,
author = {Kim, Youngsok and others},
title = {$\mu$Layer: Low Latency On-Device Inference Using Cooperative Single-Layer Acceleration and Processor-Friendly Quantization},
year = {2019},
isbn = {9781450362818},
url = {https://doi.org/10.1145/3302424.3303950},
doi = {10.1145/3302424.3303950},
articleno = {45},
numpages = {15},
location = {Dresden, Germany},
booktitle = {EuroSys'19}
}

@inproceedings{Jia2022,
author = {Jia, Fucheng and others},
title = {{CoDL: efficient CPU-GPU co-execution for deep learning inference on mobile devices}},
year = {2022},
isbn = {9781450391856},
url = {https://doi.org/10.1145/3498361.3538932},
doi = {10.1145/3498361.3538932},
pages = {209–221},
numpages = {13},
location = {Portland, Oregon},
booktitle = {MobiSys'22}
}

@inproceedings{Ling2023,
author = {Ling, Neiwen and others},
title = {{BlastNet: Exploiting Duo-Blocks for Cross-Processor Real-Time DNN Inference}},
year = {2023},
isbn = {9781450398862},
url = {https://doi.org/10.1145/3560905.3568520},
doi = {10.1145/3560905.3568520},
pages = {91–105},
numpages = {15},
booktitle = {SenSys'22}
}

@inproceedings{Jeong2022,
author = {Jeong, Joo Seong and others},
title = {Band: coordinated multi-DNN inference on heterogeneous mobile processors},
year = {2022},
isbn = {9781450391856},
url = {https://doi.org/10.1145/3498361.3538948},
doi = {10.1145/3498361.3538948},
pages = {235–247},
numpages = {13},
location = {Portland, Oregon},
booktitle = {MobiSys'22}
}

@inproceedings{Sen2025,
author = {Sen, Tanmoy and others},
title = {{Flex: Fast, Accurate DNN Inference on Low-Cost Edges Using Heterogeneous Accelerator Execution}},
year = {2025},
isbn = {9798400711961},
url = {https://doi.org/10.1145/3689031.3696067},
doi = {10.1145/3689031.3696067},
pages = {507–523},
numpages = {17},
location = {Rotterdam, Netherlands},
booktitle = {EuroSys'25}
}

@INPROCEEDINGS{Li2022:GPU-CPU,
  author={Li, Hao and others},
  booktitle={RTCSA'22}, 
  title={{Enabling Real-time AI Inference on Mobile Devices via GPU-CPU Collaborative Execution}}, 
  year={2022},
  volume={},
  number={},
  doi={10.1109/RTCSA55878.2022.00027}}

@misc{coral_edgetpu_compiler,
  title        = {{Edge TPU Compiler}},
  author       = {{Coral AI}},
  howpublished = {\url{https://www.coral.ai/docs/edgetpu/compiler}},
  note         = {Accessed: 2026-01-19}
}

@misc{onnx_graphsurgeon,
  title        = {{ONNX GraphSurgeon}},
  author       = {{NVIDIA}},
  howpublished = {\url{https://pypi.org/project/onnx-graphsurgeon/}},
  note         = {Accessed: 2026-01-20},
  year         = {2025},
}

@article{Shah:MDK,
author = {Shah, Ankit and others},
title = {{Adaptive Alert Management for Balancing Optimal Performance among Distributed CSOCs using Reinforcement Learning}},
year = {2020},
issue_date = {Jan. 2020},
publisher = {IEEE Press},
volume = {31},
number = {1},
issn = {1045-9219},
url = {https://doi.org/10.1109/TPDS.2019.2927977},
doi = {10.1109/TPDS.2019.2927977},
journal = {IEEE Trans. Parallel Distrib. Syst.},
month = jan,
pages = {16–33},
numpages = {18}
}

@article{Sun2025:SAPar,
author = {Sun, Binqi and others},
title = {SAPar: A Surrogate-Assisted DNN Partitioner for Efficient Inferences on Edge TPU Pipelines},
year = {2025},
issue_date = {September 2025},
publisher = {Association for Computing Machinery},
address = {New York, NY, USA},
volume = {24},
number = {5s},
issn = {1539-9087},
url = {https://doi.org/10.1145/3761813},
doi = {10.1145/3761813},
journal = {ACM Trans. Embed. Comput. Syst.},
month = sep,
articleno = {139},
numpages = {26}
}

@INPROCEEDINGS{Zou2024:TPUpipeline,
  author={Zou, Bohua and others},
  booktitle={MILCOM 2024}, 
  title={{A Performance Prediction-based DNN Partitioner for Edge TPU Pipelining}}, 
  year={2024},
  volume={},
  number={},
  pages={1-6},
  doi={10.1109/MILCOM61039.2024.10773756}}

@inproceedings{Yin2025:RESPECT,
author = {Yin, Jiaqi and others},
title = {{RESPECT: Reinforcement Learning Based Edge Scheduling on Pipelined Coral Edge TPUs}},
year = {2025},
isbn = {9798350323481},
url = {https://doi.org/10.1109/DAC56929.2023.10247706},
doi = {10.1109/DAC56929.2023.10247706},
pages = {1–6},
numpages = {6},
location = {San Francisco, California, United States},
booktitle = {DAC'23}
}

@misc{coral_pipeline,
  title        = {{Pipeline a Model with Multiple Edge TPUs}},
  howpublished = {\url{https://gweb-coral-full.uc.r.appspot.com/docs/edgetpu/pipeline/}},
  note         = {Accessed: 2026-01-21},
  organization = {Coral, Google}
}

@INPROCEEDINGS{Villarrubia2023,
  author={Villarrubia, Jorge and others},
  booktitle={PDP'23}, 
  title={{Improving inference time in multi-TPU systems with profiled model segmentation}}, 
  year={2023},
  volume={},
  number={},
  pages={84-91},
  doi={10.1109/PDP59025.2023.00020}}

@INPROCEEDINGS{Yin2022,
  author={Yin, Jiaqi and others},
  booktitle={SEC'22}, 
  title={{Exact Memory- and Communication-aware Scheduling of DNNs on Pipelined Edge TPUs}}, 
  year={2022},
  volume={},
  number={},
  pages={203-215},
  doi={10.1109/SEC54971.2022.00023}}

@INPROCEEDINGS{Reidy2023:TPUtransformer,
  author={Reidy, Brendan and others},
  booktitle={RTAS'23}, 
  title={{Work in Progress: Real-time Transformer Inference on Edge AI Accelerators}}, 
  year={2023},
  volume={},
  number={},
  pages={341-344},
  doi={10.1109/RTAS58335.2023.00036}}

@misc{tflite_converter,
  title        = {{tf.lite.TFLiteConverter}},
  howpublished = {\url{https://www.tensorflow.org/api_docs/python/tf/lite/TFLiteConverter}},
  note         = {Accessed: 2026-01-21},
  organization = {Google},
  year         = {2024},
  description  = {TensorFlow Lite Converter API for converting TensorFlow models to TensorFlow Lite format.}
}

@article{Wang2025:OnDeviceAISuvery,
author = {Wang, Xubin and others},
title = {{Empowering Edge Intelligence: A Comprehensive Survey on On-Device AI Models}},
year = {2025},
volume = {57},
number = {9},
issn = {0360-0300},
journal = {ACM Comput. Surv.},
month = apr,
articleno = {228},
numpages = {39},
}

@inproceedings{Dong2025:Real-Offload,
author = {Dong, Qifei and Xu, Jingao and Pillai, Padmanabhan and Satyanarayanan, Mahadev},
title = {{Does Accurate Real-Time AI Need Edge Offload?}},
year = {2025},
isbn = {9798400722387},
booktitle = {{SEC '25}},
articleno = {20},
numpages = {18}
}

@INPROCEEDINGS{Abdelzaher2018:IoBT,
  author={Abdelzaher, Tarek and others},
  booktitle={ICDCS}, 
  title={{Will Distributed Computing Revolutionize Peace? The Emergence of Battlefield IoT}}, 
  year={2018},
  volume={},
  number={},
  pages={1129-1138}}

@misc{raspberrypi_ai_hat,
  title        = {{Raspberry Pi AI HAT+}},
  howpublished = {\url{https://www.raspberrypi.com/products/ai-hat/}},
  note         = {Accessed: 2026-01-23},
  organization = {Raspberry Pi Ltd.}
}
\bibliographystyle{ieeetr}     
\end{document}